\documentclass[conference]{IEEEtran}
\IEEEoverridecommandlockouts
\usepackage{color}
\usepackage{cite}
\usepackage{amsmath,amssymb,amsfonts}
\usepackage{algorithmic}
\usepackage{graphicx}
\usepackage{textcomp}
\usepackage{xcolor}
\def\BibTeX{{\rm B\kern-.05em{\sc i\kern-.025em b}\kern-.08em
    T\kern-.1667em\lower.7ex\hbox{E}\kern-.125emX}}
\usepackage[T1]{fontenc}

\usepackage{ragged2e}
\begin{document}

\title{Hyper: Distributed Cloud Processing for Large-Scale Deep Learning Tasks 
\thanks{Funded by Snark AI, Inc.}
}

\author{\IEEEauthorblockN{Davit Buniatyan}
\IEEEauthorblockA{\textit{Snark AI, Menlo Park, CA 94025} \\
\IEEEauthorblockA{\textit{Princeton University, Princeton, NJ 08544}}
davit@snark.ai}
}

\maketitle

\begin{abstract}
Training and deploying deep learning models in real-world applications require processing large amounts of data. This is a challenging task when the amount of data grows to a hundred terabytes, or even, petabyte-scale.  We introduce a hybrid distributed cloud framework with a unified view to multiple clouds and an on-premise infrastructure for processing tasks using both CPU and GPU compute instances at scale. The system implements a distributed file system and failure-tolerant task processing scheduler, independent of the language and Deep Learning framework used. It allows to utilize unstable cheap resources on the cloud to significantly reduce costs. We demonstrate the scalability of the framework on running pre-processing, distributed training, hyperparameter search and large-scale inference tasks utilizing 10,000 CPU cores and 300 GPU instances with overall processing power of 30 petaflops.\footnote{ Demo available at https://lab.snark.ai}\footnote{ Documentation available at https://docs.snark.ai }
\end{abstract}


\section{Introduction}

Deep Learning (DL) based models have outperformed manual feature engineered algorithms in a wide range of domains including computer vision, natural language processing, audio processing \cite{krizhevsky2012imagenet, lecun2015deep, goodfellow2016deep, zhang2015character, mikolov2013efficient, bahdanau2014neural, redmon2016you}.  The amount of labelled data required for training production-ready models achieves a terabyte-scale \cite{imagenet_cvpr09, tencent-ml-images-2019}. The unlabelled data to execute those models reaches a petabyte-scale \iftrue \cite{zheng2018complete, commoncrawl, akiyama2019first} \fi. Such computational resources are on-demand available on clouds such as Amazon Web Services (AWS), Google Cloud Platform (GCP) or Azure. 

Modern Deep Learning frameworks such as PyTorch \cite{pytorch}, Tensorflow \cite{tensorflow}, MXNet \cite{mxnet}, Theano \cite{theano} and Caffe \cite{caffe}  are well-positioned for training and deploying models on single multicore machines with multiple GPUs or TPUs. Training state-of-the-art models on terascale data often takes weeks or months to converge. To make training faster, DL frameworks such as PyTorch or Tensorflow recently introduced synchronous and asynchronous distributed training methods to achieve an almost linear speedup with respect to the number of nodes.

%Additionally, one can also choose third-party distributed training frameworks such as Horovod \cite{sergeev2018horovod} with MPI based implementation \cite{mpi}. If the network between compute nodes is well set, for example, using InfiniBand connection \cite{liu2004high}, it is possible to achieve almost linear speedup with respect to the number of nodes available. 

As the number of nodes in a distributed cluster grows, problems such as provisioning, orchestrating, fault-tolerance, distribution data management, task planning, and execution arises. For executing Big Data workloads, several widely accepted synchronous parallel processing systems have been introduced such as MapReduce\cite{dean2008mapreduce}, Apache Spark\cite{zaharia2010spark} and Dryad \cite{isard2007dryad}.  Additionally, task-parallel frameworks are getting increased usage such as Dask\cite{rocklin2015dask} or CIEL\cite{murray2011ciel}. They provide fine-grained task management. These frameworks are very efficient for ETL tasks but lack native deep learning support. A recent introduction to the family is Ray, which implements a dynamic task-parallel framework that is suited for deep learning and reinforcement learning \cite{moritz2018ray}.

To manage High-Performance Computing (HPC) workloads, container technology has recently become well suite choice for packaging environment libraries \cite{xavier2013performance}. Frameworks such as Kubernetes enable massive distribution of stateless applications packaged in a container on a cluster of nodes in fault-tolerant way \cite{bernstein2014containers}. However, it still lacks an efficiently distributed data management unit, workflow scheduling system, and support for stateful applications such as model training. Packages such as KubeFlow and Argo attempt to extend Kubernetes to support the implementation of machine learning pipelines \cite{huang2018kubebench}. 

For data-intensive workloads, a variety of distributed file storage systems have been introduced \cite{satyanarayanan1990survey} such as NFS \cite{nfs} or HDFS \cite{hdfs}. NFS-based file systems significantly decrease multi-read speed and lower bound the computing speed. They often do not scale on multi-write scenarios. There is always a trade-off between latency and scalability. For web-based applications, cloud providers offer an object storage solution that has high scalability but suffers from low-latency memory-intensive operations. 

\begin{figure*}[h]
    \centering
    \includegraphics[width=\textwidth]{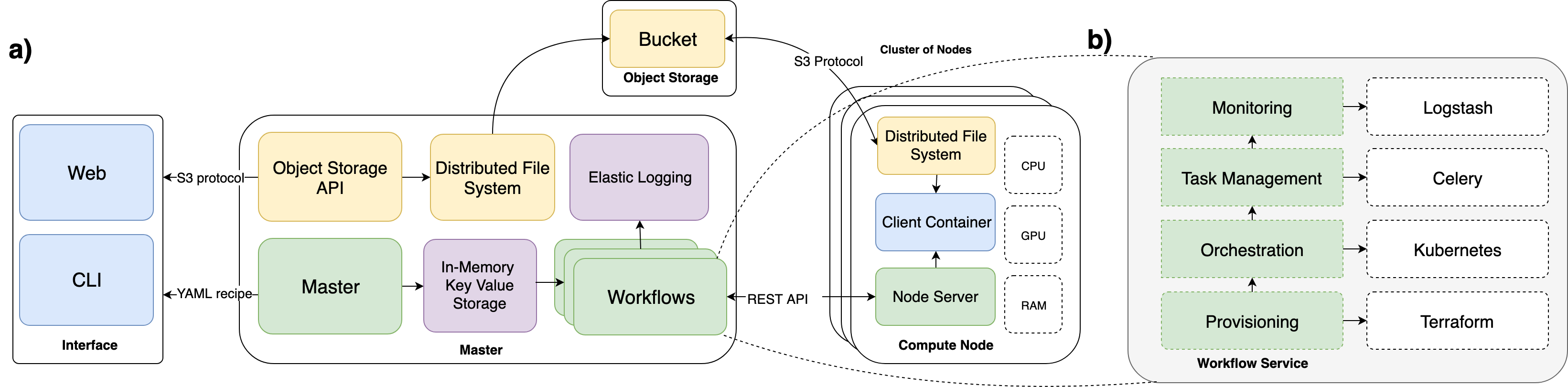}
    \caption{ System Architecture a) Interface uploads the training data, source code and YAML recipe to the Master Node. Source files are chunked and uploaded to Object Storage. The Recipe is parsed to create a computational graph in in-memory Key-Value Storage. For each workflow, the cluster is created in the cloud. Each computational node has a Node Server that handles the management of the node and executing client containers. b) Workflow has four main stages. Provisioning the infrastructure, orchestrating nodes, executing tasks and monitoring workers.}
    \label{fig:architecture}
\end{figure*}

We introduce a hybrid distributed cloud framework with a unified view of multiple clouds and on-premise infrastructure for processing tasks. We have made the following contributions
\begin{itemize}
  
  \item We design and build a distributed framework that unifies preprocessing, training, hyperparameter search and deploying large scale applications on the cloud.
  \item To achieve scalability we design and deploy a distributed file system that has near-zero delay for deep learning jobs in comparison to downloading the data locally on the machine with similar performance.
  \item Provide fault-tolerance with respect to computational node failures and support utilization of unstable cheap resources on the cloud. 
  \item Abstract away cloud infrastructure setup and provisioning from the user with native cloud integration. 
\end{itemize}

\section {Computational Model}

In this section, we discuss design decisions made for the system and user interface to specify the computation workload.
\subsection{Computational Workflows}
\textit{Workflow} is a directed acyclic graph consisting of \textit{Experiment} nodes and their dependency as edges. Single $Experiment$ contains multiple \textit{Tasks}. $Tasks$ within the same experiment execute the same command with different arguments. Arguments can be templated for efficient parameter space definitions. Each \textit{Experiment} has an associated container that is deployed on all computational workers.

\textit{Task} is the execution unit, which encapsulates a process. Each \textit{Task} has assigned \textit{Node}, which represents the computation worker. Single \textit{Node} might execute multiple \textit{Tasks}. The number of nodes available corresponds to the number of workers inside the cluster. 

\subsection{Interface}

$Workflows$ are specified using code-as-infrastructure interface defined in YAML recipes as seen in Fig. \ref{interface:3}. The recipe is parsed by the server and translated into a directed acyclic graph of experiments. The interface lets users specify the environment, hardware settings, number of workers, parameters and parameterized commands. The user can interface the system through CLI or Web UI.

\subsection{Parameters}
The user can specify the list of parameters that can be inserted into command arguments during execution. Parameters can be sampled from a discrete class or continuous range. To compute parameters for each \textit{Task}, the algorithm generates the Cartesian product of all discrete parameters and samples from the set $n$ times with minimal repetition. $n$ is defined to be the number of samples from a recipe. Then, it samples $n$ times from each continuous parameter range and randomly matches with discrete sampled parameters. This is necessary to support both hyper-parameter search and inference with grid iterators.

\medskip
\section {System Overview}

The system receives data, chunks it and stores it in object storage. The recipe is submitted to deploy the deep learning workflow on a cluster of nodes, which mounts the distributed file system.

\subsection{Distributed Data Management}
%Object storage is considered to be a highly scalable and appropriate choice for storing large scale amounts of data, however not very efficient for memory-intensive HPC jobs. An alternative offered by TensorFlow \cite{tensorflow} is to store the data on object storage in tfrecord files and use programmatic access to fetch the data continuously in the background. This implementation works well but is limited only to a single Deep Learning package and requires an initial step to transform datasets into the required form. When one has specific constraints about data model, for example, it is a large volumetric data as seen in ChunkFlow \cite{wu2019chunkflow}, then you can chunk the data into small pieces and upload it to the object-store. Then, to run inference one can asynchronously fetch objects. This assumption may work well for deep learning tasks in general because of tensor input and output paradigm, however, not every DL application has preferred storage as tensors. For example, in natural language processing, it is more efficient to store the text in characters rather than one-hot encoding of words. 

In order to be framework agnostic and require no further modification of the client program, we chunk the file system itself and store it in object storage provided by the cloud vendor (e.g. AWS S3). The system implementation is similar to closed source ObjectiveFS \cite{ofs} and tuned for deep learning tasks. The distributed file-system wraps POSIX API and acts as a middle layer with chunking, caching and state synchronization mechanisms across all nodes.
When the program queries the file system for a specific file, the integration layer checks which chunk contains the file to download. In the next query, the file system can check if the existing chunk contains the next required file before fetching it from the cloud. Within the program's context, files that are stored in remote chunked object storage appear to be local files. Any deep learning application without further modification will take advantage of the highly scalable storage. 

Deep learning frameworks such as PyTorch and TensorFlow natively support asynchronous data fetching from the local storage to the GPU using data loaders. Often the deep learning training iteration is bounded by the compute cycles on GPUs. If one combines the distributed remote storage and asynchronous data fetching, the training speed is almost the same as if the data was stored locally on the machine with respect to the following constraints.

\begin{figure}[h]
  \centering
  \includegraphics[width=250px]{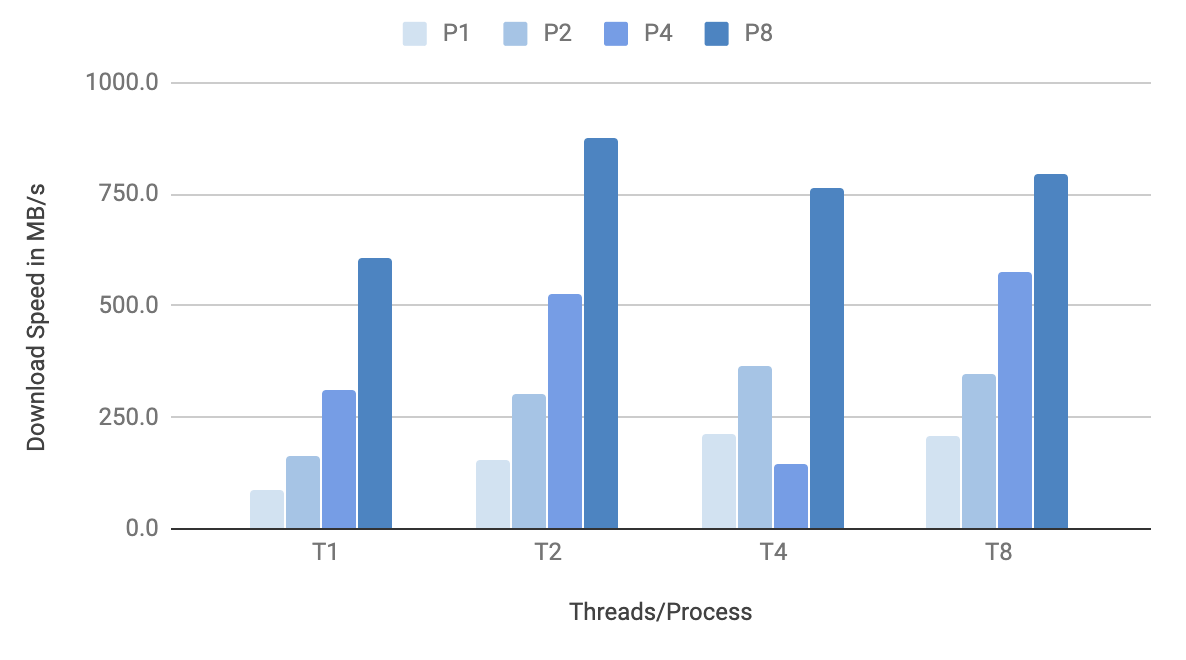}
  \caption{Hyper File System on a single machine (AWS p3.2xlarge) can achieve up to 875 MB/s download speed with multithreading $T$ and multiprocessing $P$ enabled.  The batch size was chosen smaller for large models to fit in the GPU RAM. }
  \label{fig:ChunkPar}
\end{figure}

\begin{itemize}
  \item Network locality - We assume that computational nodes that access the data are physically located near object storage servers.
  \item Chunksize - well chosen with respect to latency to maximize the throughput as shown in Fig. \ref{fig:ChunkPar}. It should be in the range 12-100MB. 
\end{itemize}

The suggested file system can leverage the scalability of the object storage and provide data access to the cluster of nodes with almost native speed for deep learning jobs as shown in Fig. \ref{fig:HyperFS}.

\subsection{Cloud Infrastructure}
Distributed frameworks such as Spark, provide a functional ecosystem for computational tasks. End-users should only specify the program and apply it to large datasets. User is limited to only using supported libraries abstracted in MapReduce framework. State-of-the-art libraries are out of reach.  

%for building decision trees \cite{chen2016xgboost, lightgbm} or processing natural language \cite{Gardner2017AllenNLP, Manning14thestanford}.  %\cite{meng2016mllib} 

\textbf{Provisioning}: When constraints of the system does allow arbitrary package support, then the whole environment and necessary packages should be transported to the computational node. We use container technology to bake the necessary libraries. Compute nodes need to have docker support to execute arbitrary containers and Nvidia CUDA \cite{kirk2007nvidia} libraries for processing deep learning operations on the GPU.  

\textbf{Orchestration}: Due to its generic feature of container technology, the Virtual Machine (VM) images necessary to run containers can be based on any Unix based operating systems including CoreOS, Ubuntu and CentOS. VM images are built only once and stored in the cloud. It acts as a proxy to execute custom specified containers. The user can specify the container from the public repository. After cloud instances are provisioned, each instance downloads a client container. This mechanism allows supporting any framework, library or package without constraints. We also cache frequently used containers such as Tensorflow, Pytorch, Jupyter directly inside VM images to reduce loading time. In addition to custom docker management, the system can offload container orchestration to managed Kubernetes \cite{bernstein2014containers}.

\textbf{Networking}: For cloud infrastructure orchestration, we use Terraform, which provides code-as-infrastructure language for defining cloud resources. For each job execution, the system specifies a Virtual Private Network with Internet Gateway. It makes cluster nodes accessible inside the network for use cases that require state synchronization across nodes such as during distributed training. Alternatively one can use object storage as a parameter server to store the model without networking setup. 

\subsection{Implementation}

As shown in Fig. \ref{fig:architecture}, the architecture of the distributed framework consists of main components: Interface, Master and Node. Master is responsible for receiving the recipe of the pipeline, parsing and creating workflow objects including experiments and tasks. The objects are stored in-memory key-value cache Redis. As a backup alternative, the system stores the state into DynamoDB. Then, the master starts a new workflow service as an adjacent container to orchestrate and schedule tasks. During the orchestration process, each compute worker runs a node server that listens to commands executed by the workflow manager. Each node starts to pull the client specified container and mount the distributed file system. Once the node is ready, the workflow manager can execute the client's specified commands. 

There are three types of logs that are collected into Elastic Logstash: client application logs, CPU/GPU utilization logs and operating system logs \iftrue \cite{gormley2015elasticsearch} \fi. 

In addition to the scheduling system, we deploy our own object storage layer for providing S3-like API \iftrue \cite{sivasubramanian2012amazon}\fi to client interface. When the data is uploaded to Minio server \iftrue\cite{minio}\fi, files get chunked and stored on the distributed file system. 
Celery is used for asynchronous task management. The task management system is similar to Apache Airflow \iftrue \cite{airflow} \fi and other workload management systems such as Splunk. It is different from frameworks such as Dask \cite{rocklin2015dask} \iftrue, CIEL \cite{murray2011ciel} \fi or Ray \cite{moritz2018ray} in terms of execution granularity. 

\subsection{Fault tolerance}

In order to optimize cloud resource allocation, cloud vendors provide spots or preemptible instances. Those instances are usually 2 or 3 times cheaper but can be terminated anytime depending on the demand and the price per hour bid. For stateful long-lasting jobs, cost optimization is an attractive option, however, it requires additional compute logic implementation to recover the state.

Since the system already provides a distributed file system backed by remote object storage and a scheduling system, it becomes straightforward to implement a fault-tolerant system that can handle instability. When a node fails, the task with exact command arguments gets rescheduled on a different node. In training use cases, modern deep learning frameworks provide an easy interface to store and retrieve model states. Hence, the training can be continued without any additional code modifications. 

\medskip

\section{Evaluation}
%In our evaluation, we study the following questions 
%\begin{itemize}
%  \item How Hyper meets data preprocessing achieved?
%  \item How does a distributed setup scale with the number of machines?
%  \item How well can the interface extend to the custom package hyperparameter search scenario?
%  \item What advantages does Hyper provide for large scale inference tasks?
%\end{itemize}

To evaluate the system we run pre-processing, distributed training, hyper-parameter search and large scale inference using AWS CPU M5 family and GPU P3 family compute instances. We use AWS S3 to store file system chunks.

\subsection{Preprocessing}

To test the scalability of the system for ETL tasks, we set up a preprocessing experiment. 100 million text files from commoncrawl \iftrue \cite{commoncrawl} \fi dataset are uploaded to the distributed storage. The amount of data achieves ~10TB. We specify the infrastructure to spin up 110 instances each with 96 CPU cores. The processing script takes 100,000 text files and transforms them into tfrecord files. During the transformation, spaCy \iftrue \cite{spacy} \fi package is used for filtering, tokenizing and splitting paragraphs. We also enable spot instances to reduce costs and test fault tolerance. 

\subsection{Distributed Training}

We defined a recipe for training object recognition model YoloV3\iftrue \cite{redmon2018yolov3} using PyTorch \cite{pytorch} using Horovod \cite{sergeev2018horovod} \fi. Then, we uploaded COCO dataset \cite{lin2014microsoft} to the storage. The training script reads all images and labels from the defined path.  Furthermore, parameters such as how many epochs, learning rate and what input image size are parameterized in the recipe. We also parametrize model-specific parameters such as a total number of classes,  confidence threshold, nms threshold and iou threshold. 

Nvidia K80 GPUs are slow to train the model. With a single line configuration change, we deployed the training on Nvidia V100 GPUs with spot enabled instances to reduce costs. The batch size for the training was accordingly modified. The cost would be \$8.48/h instead of \$0.95/h, but the training is 50x faster with 6x efficiency gain. We modified hyper-parameters and started another experiment with zero effort. 

\begin{figure}[h]
  \centering
  \includegraphics[width=250px]{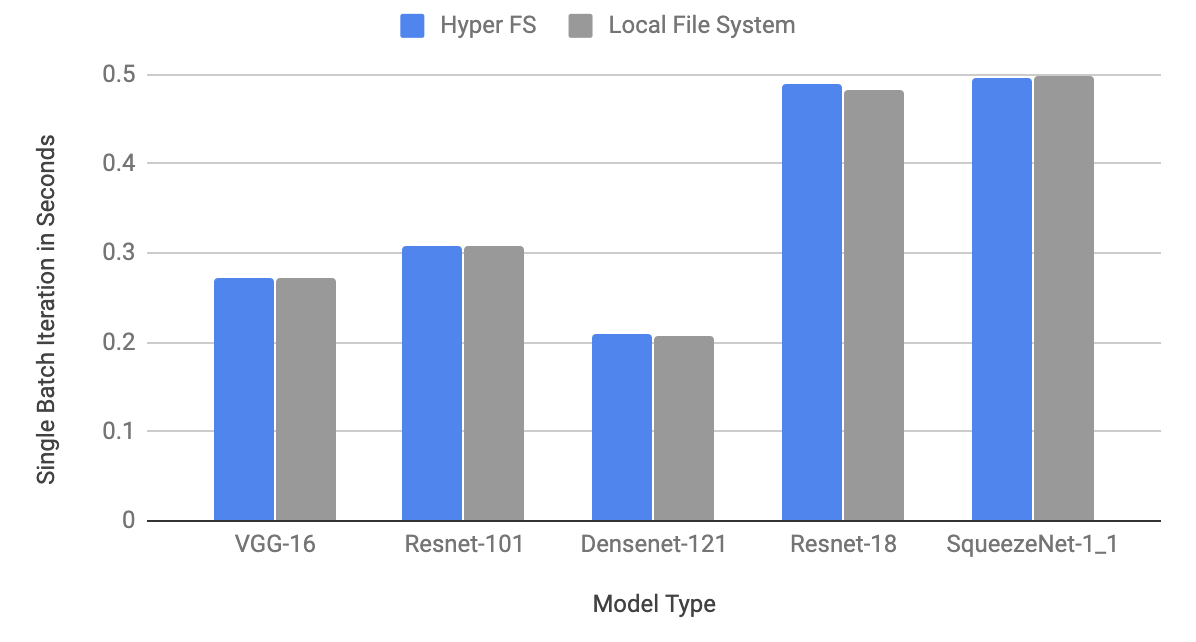}
  \caption{Streaming data through Hyper File System while training a deep learning model is equivalent to reading data from the local file system}
  \label{fig:HyperFS}
\end{figure}

We also provide benchmarking of data streaming against storing data in local files while training data-intensive computer vision models as shown in Fig. \ref{fig:HyperFS} and Fig. \ref{fig:HyperFS}. 

\subsection{Hyperparameter search}

Gradient boosting machine, XGBoost\cite{chen2016xgboost} or LightGBM \cite{lightgbm}, is one of the best off-the-shelf machine learning solvers for tabular data, however training those models can be computationally very heavy and have a lot of parameters to tune. There are 12 parameters to tune for the tree booster. If you try 2 choices for each one, there will be 4096 different combinations. If each training takes 10 mins to complete, trying out all those 4096 combinations sequentially would take 28.4 days. Using our system, we made the experiments run in 10 minutes by linearly increasing the cluster size without source code modification. 

%Implementations such as XGBoost \cite{chen2016xgboost} and LightGBM \cite{lightgbm} have been one of the essential components for winning solutions of machine learning competitions.

%XGBoost is an iterative algorithms that construct a tree in each step of the computation. Each tree construction takes time. To avoid overfitting, it is usually preferable to run XGBoost with smaller step sizes with a larger number of iterations. 

%K-fold Cross-Validation (CV). To evaluate your model, it is a common practice to run the training on subsampled training data for K times and compute test scores on the left-out training data. It makes the overall time K times the single training time unless you parallelize the CV process. 

\begin{figure}[h]
  \centering
  \includegraphics[width=250px]{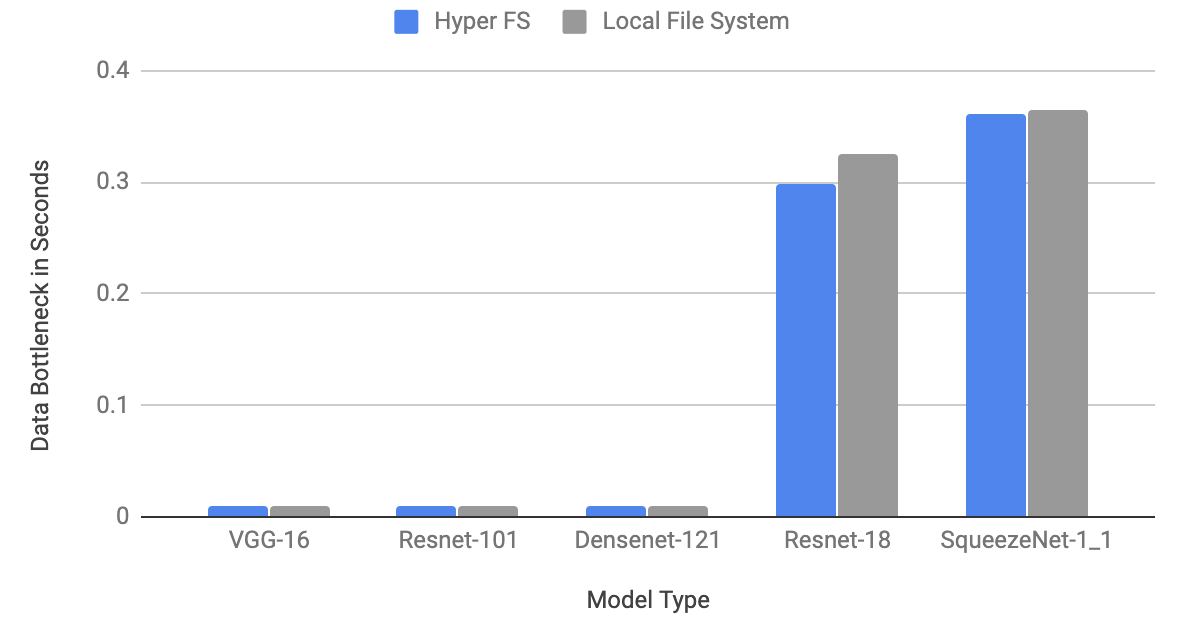}
  \caption{In the asynchronous data loading figure, the first three models (VGG, Resnet101, and DenseNet) have no data bottleneck. The benchmarks have been produced on AWS using a p3.2xlarge machine with V100 GPU. The data is stored on S3 within the same region.}
  \label{fig:Databottleneck}
\end{figure}

\subsection{Large-Scale Inference}
For production-ready processing, we upload an ImageNet dataset \iftrue \cite{imagenet_cvpr09} \fi and split it into 300 folders. Each folder contains 1500 images. Using Hyper interface, we easily parallelized the inference execution of Yolo model to 300 GPU instances with the overall processing of 2 petaflops. 

\iftrue
\section {Discussion}

Building Hyper was inspired by processing a petascale amount of brain images for connectomics \cite{zheng2018complete}. At the time, Map-Reduce based frameworks such as Spark and directed graph processing frameworks such as Dask or Ray have not been efficient in executing deep learning computes. Initially, we used AirFlow \iftrue \cite{airflow} \fi, then used framework ChunkFlow \cite{wu2019chunkflow}. Hyper is another generation of cloud-agnostic, data-agnostic distributed deep learning computed with scale. Since Hyper was released hundreds of people have used the platform and several companies have used it in production. Here we discuss some of the early feedback and limitations of the software.

\textbf{Feedback} When  Hyper is used in production, it completely encapsulates infrastructure and is flexible enough to customize projects under user needs. The user only needs to upload the data and the source code to the distributed file storage through CLI or Web UI. Then submit the infrastructure as minimal code to describe the compute workflow. It provides interface to log results of hyperparameter search. 

\textbf{Ecosystem} Hyper supports any open source machine learning libraries and frameworks. We can offer any pre-built algorithms and models that suit different business cases.

\textbf{On-prem Deployment} They system can deploy the software to any on-premise infrastructure with Kubernetes support.

\textbf{Built-in Modeling Algorithms} We support cutting-edge pre-built algorithms and models for a wide range of industry use-cases such as image recognition, speech to text, text to speech, anomaly detection, recommendation and personalization, and others.

\textbf{Unlimited Persistent Data Lake} Scalable and cost-efficient persistent central Data Lake that merges data from all of your different data sources. A single source of truth for a complete view of the datasets. 10x cheaper on the storage system compared to AWS EFS expenses with almost all native speed of data read/write. 

\textbf{Spot/Pre-emptible Instance Cost Savings} AWS and GCP offer dynamic bidding for instances which are usually 3x cheaper than normal instances but also can be stopped at any time. Spot instance management layer makes it very easy to use spot instances and also enjoy the cost-saving.

\textbf{Full lifecycle ML pipeline} Drag and drop machine learning modules that include data ingestion modules, data ETL modules, built-in model training algorithm modules, hyper-parameter search modules, and model deployment modules. Combine the modules into a pipeline for a full lifecycle of your machine learning task.

%\begin{figure}[h]
%\begin{minted}{yaml}
%experiments:       # List of experiments
%  mnist:           # Name the experiment
%    framework: pytorch # Or specify Docker
%    parameters:    # Define parameters
%     lr:
%       range: 0.1-0.3
%       sampling: uniform
%    samples: 1000  # Number of samples
%    workers: 100   # Number of workers
%    hardware:      # Set hardware reqs.
%      gpu: k80
%    command:       # Command executed
%      - python train.py --lr {{lr}}
%\end{minted}
%\caption{Example recipe that does hyperparameter search. The user specifies the DL framework or docker container URL, defines parameters, number of samples to be drawn from parameter space, number of workers inside a cluster,  the hardware definition of the computational node and the parameterized command. The recipe is uploaded to REST endpoint and executed on the cloud.}
%\label{interface:3}
%\end{figure}

\textbf{Limitations} Current limitations include low interactivity of the workflow. Once the workflow has been defined. Users are not able to modify the execution unless terminated or start a new one. Compared to Ray and Dask, Hyper modifies the surroundings of the application-defined by deep learning researcher - and has minimal impact on the code itself. Even though it brings several advantages for end-users, it also provides strict limitations on computing optimization and efficiency. Optimizations such as GPU full capacity utilization and inter-network communications are still handled by the user. 
\fi

\section {Conclusion}

Hyper provides a unified view of multiple clouds and on-premise infrastructure without requiring a team of DevOps engineers to save AI/ML-Ops time. It provides cloud cost savings and transparent compute resource utilization tracking. Hyper Storage solutions are significantly better for data-intensive deep learning tasks compared to cloud-native NFS-like offers and much more cost-efficient since they are backed by object storage.

Hyper enables data scientists to be highly efficient in machine learning and deep learning. It provides a framework for running experiments, collecting logs and comparing models including one-click Jupyter notebooks or Tensorboard graphs using Web UI or CLI. Data scientists can plug-and-play state-of-the-art deep learning models in Computer Vision, NLP, and other domains to kickstart their project. They can execute large-scale distributed training or batch processing jobs through a very simple interface without knowing about the infrastructure and define continuous workflows for automatic model training, validation, benchmarking and deployment.

Future work includes the development of various Bayesian optimization algorithms for hyper-parameter tuning of models, seamless integration with Kubernetes and interactive workflows. 

%  inside workflow similar to RayTune \cite{liaw2018tune},

% Sample Figure ------------------------------------------------------------
%\begin{figure}[h]
%\label{sample-figure}
%\begin{center}
%\includegraphics[width=5cm]{sample.eps}
%\caption{Sample figure}
%\end{center}
%\end{figure}

% Sample Table ------------------------------------------------------------
%\begin{table}[h]
%\renewcommand{\arraystretch}{1.2}
%\caption{Sample table}
%\vspace{1mm}
%\label{sample-table}
%\begin{center}
%\begin{tabular}{|c|c|c|}
%\hline
%Title 1 & Title 2  & Title 3\\
%\hline
%item 1,1 & item 1,2  & item 1,3\\
%\hline
%item 2,1 & item 2,2  & item 2,3\\
%\hline
%item 3,1 & item 3,2  & item 3,3\\
%\hline
%\end{tabular}
%\end{center}
%\end{table}

% Sample Theorem ----------------------------------------------------------
%\begin{theorem}
%This is the sample theorem...
%\end{theorem}
%
%\proof The proof of the theorem.
%\endproof

\iftrue \section{Acknowledgement}
The authors would like to thank Jason Ge, Sergiy Popovych, Vazgen Hakobjanyan, Gevorg Soghomonyan, Sebouh Der Kiureghian, Gegham Jivanyan, Gevorg Abovyan, Nazar Balyan. Also would like to thank AWS for providing cloud resources for experiments. The project was funded and supported by Snark AI, Inc. \fi

% References -------------------------------------------------------------
\bibliographystyle{plain} 
\bibliography{refs}

\end{document}